\documentclass[%
 reprint,
 amsmath,amssymb,
 prl,
]{revtex4-2}

\usepackage{graphicx}% Include figure files
\usepackage{dcolumn}% Align table columns on decimal point
\usepackage{bm}% bold math
\usepackage[colorlinks=true, linkcolor=blue,
  citecolor=olive,urlcolor=blue]{hyperref} % add hypertext capabilities
\usepackage[dvipsnames]{xcolor}

\usepackage{soul}

\DeclareMathOperator{\erf}{erf}
\DeclareMathOperator{\ph}{ph}
\DeclareMathOperator{\sgn}{sgn}
\DeclareMathOperator{\Imi}{Im}

\begin{document}

\preprint{APS/123-QED}

\title{Time Evolution of a Decaying Quantum State:\\Evaluation and Onset of Non-Exponential Decay}% Force line breaks with \\
%\thanks{A footnote to the article title}%

\author{Markus N\"oth}
 \email{noeth@math.lmu.de}%Lines break automatically or can be forced with \\
\affiliation{Mathematisches Institut, Ludwig-Maximilians-Universit\"{a}t M\"{u}nchen, Theresienstr. 39, D-80333 M\"{u}nchen, Germany}%

\date{\today}% It is always \today, today,
             %  but any date may be explicitly specified

\begin{abstract}
The method developed by van Dijk, Nogami and Toyama \cite{vandijk1,vandijk2,vandijk3,vandijk4} for obtaining the time-evolved wave function of a decaying quantum system is generalized to potentials and initial wave functions of non-compact support. The long time
asymptotic behavior is extracted and employed to predict the timescale for the onset of non-exponential decay. The method is illustrated with a Gaussian
initial wave function leaking through Eckart's potential barrier on the halfline. 
%\begin{description}
%\item[Usage]
%\item[Structure]
%\end{description}
\end{abstract}

%\keywords{Suggested keywords}%Use showkeys class option if keyword
                              %display desired
\maketitle

%\tableofcontents

%\section{\label{sec:level1}First-level heading:\protect\\ The line
%break was forced \lowercase{via} \textbackslash\textbackslash}
\section{Introduction}

Quantum tunneling, the phenomenon of a particle traversing a potential barrier
higher than its kinetic energy, has received a great deal of attention since the
early works of Gamow \cite{gamow}. Despite the time and effort spent on this
issue (see \cite{hartman1962tunneling, muga2007time, theoryoftunneling,
hauge1989tunneling} for important contributions to the discussion and reviews
and \cite{steinberg2020measurement} for a recent experiment), there still remain
open questions about tunneling. One reason for this is that while tunneling is
fundamentally a process in time, much of the analysis has been restricted to
static aspects of the relevant setups. While important properties of tunneling
systems such as exponential decay rates describing a decaying system for
intermediate times can be found in such an analysis, a more thorough
understanding seems desirable.

For this reason studying the time dependent wave function describing the
tunneling particle is essential to complete the picture. For the setup of a
decaying quantum state a great deal of progress in this direction has been
achieved by van Dijk, Nogami and Toyama
\cite{vandijk1,vandijk2,vandijk3,vandijk4}. For a precursor to the technique see
\cite{garcia1996survival}. They found a representation of the time dependent
wave function well suited for studying early, intermediate and late times. This
makes it well suited for analytic as well as numerical endeavors. The method was
only applied to compactly supported potentials and initial wave functions. The
compact support of the initial wave function played an important role, as will
be explained in the next section. In this article I extend this method and apply
it to a Gaussian initial wave packet leaking out of Eckart's potential  barrier;
neither of which has a compact support. The necessary modification of the
technique is introduced first.

Following a reminder of the tricks employed and the generalization of the
present work I analyze the late-time asymptotics. While the power-law for the
decay of the wave function
\begin{equation}\label{asymptotics}
|\psi(r,t)|\overset{t\rightarrow \infty}{\rightarrow} \frac{|\psi_\infty(r)|}{t^{3/2}},
\end{equation}
in the generic case is well-known, I derive a simple and fairly general
expression for the constant involved cf. \cite{jensen1979spectral}. Furthermore,
an estimate of the time \(t_{\text{alg}}\) after which the transition from
exponential to algebraic decay takes place is given in \eqref{t_late}. This
result should be compared to the one for the survival amplitude by
\cite{peres1980nonexponential}. For a complicated barrier it may not be feasible
to apply the full technique while the expressions for \(|\psi_\infty(r)|\) and
\(t_{\text{alg}}\) can be evaluated more easily and may therefore be of greater
importance when comparing with experiments.

\section{Recap and Extension}
\subsection{Main Argument}
We consider the Schrödinger equation (setting \(\hbar=2m=1\))
\begin{equation}\label{schroed general}
    i \partial_t \psi = -\partial_r^2 \psi + V \psi,
\end{equation}
for the scalar wave function \(\psi\) subject to the potential \(V\) and
boundary condition \(\psi(0,t)=0\) for all \(t\), chosen such that there
are no bound states \footnote{See \cite{vandijk3} for how to modify the
technique in the presence of bound states}. This theoretical setup may describe
the radial part of a system describing alpha-decay with zero angular
momentum \cite{microscopic_alpha} or it may describe the motion of an ion along
the axis of a quantum optical trap inducing a back wall \cite{das2019arrival,
das2019exotic} and an extra trapping potential \(V\).  We will make use of the
Jost solutions \(f(k,r)\) of \eqref{schroed general} which are the generalized
eigenfunctions of the Hamiltonian of \eqref{schroed general} of energy \(k^2\)
that satisfy 
\begin{equation}
    \lim_{r\rightarrow \infty}e^{-ik r}f(k,r)=1.
\end{equation}
Jost solutions exist for potentials \(V\) such that \(\int_0^\infty dr\, r
|V(r)|<\infty\). For fixed real \(r\) the Jost solution \(f(k,r)\) is analytic
on \(\{\Imi(k)\ge 0\wedge k\neq 0\}\subset \mathbb{C}\) and fulfills
\(\overline{f(k,r)}=f(-\overline{k},r)\) \cite[Thm XI.57]{reed1979scattering}.
Furthermore, I assume that 0 is not an eigenvalue or resonance of the
Hamiltonian involved, which is typically true.

The generalized eigenfunction \(u(k,r)\) of the same energy satisfying the
proper boundary condition at the origin and the additional normalization 
\begin{equation}
    u(k,0)=0, \quad u'(k,0)=1,
\end{equation}
where the prime denotes a derivative with respect to the 
second argument, is given by 
\begin{equation}\label{u in terms of f}
    u(k,r)=\big( f(k,0) f(-k,r)-f(-k,0) f(k,r)\big) \frac{1}{2i k}.
\end{equation}
Furthermore, \(u(k,r)\) is real for real \(r\) and \(k\) and even in \(k\) and
for \(r\) fixed \(u(k,r)\) is an entire function of \(k\) \cite[Thm
XI.56]{reed1979scattering}.  The solutions \(u\) form a complete set and are
normalized to \cite{vandijk1}
\begin{equation}\label{normalization}
    \int_0^\infty \overline{u(k,r)} u(k',r) dr = \frac{\pi}{2 k^2}|f(k,0)|^2 \delta(k-k'),
\end{equation}
for \(k,k'>0\).  So we may expand any solution \(\psi(x,t)\) of equation 
\eqref{schroed general}:
\begin{align}\label{psi_integral}
\psi(r,t)=\frac{2}{\pi} \int_0^\infty \frac{k^2}{f(k,0)f(-k,0)} C(k)e^{-i k^2 t}u(k,r) dk,
\end{align}
with 
\begin{equation}
    C(k)=\int_0^\infty u(k,r) \psi(r,0) dr,
\end{equation}
where I eliminated complex conjugates by the above mentioned properties of \(u\)
and \(f\). Because of the symmetry properties of the integrand I split \(u\) up
as in equation \eqref{u in terms of f} to obtain 
\begin{equation}\label{follow vd}
    \psi(r,t)=\int_{-\infty}^\infty e^{-i k^2 t + ikr} \frac{C(k)}{i \pi}  \frac{k\, e^{-ik r} f(-k, r)}{f(-k, 0)}dk,
\end{equation}
where I added an additional plane wave so that the last factor approaches a
finite limit for \(r\rightarrow \infty\). Next I introduce an auxiliary factor
\(h(k)\)
\begin{equation}\notag
    \psi(r,t)=\frac{1}{i \pi}\int_{-\infty}^\infty  e^{-i k^2 t + ikr}h(k) \overbrace{\frac{k \,C(k)e^{-ik r} f(-k, r)}{h(k) f(-k, 0) }}^{=:g(k,r)}dk,
\end{equation}
to achieve the following properties:
\begin{itemize}
	\item The fraction \(g(k,r)\) is still meromorphic in \(k\).
    \item The expansion of \(g(k,r)\) by the Mittag-Leffler \cite{jeffreys1999methods} theorem is of the form
    \begin{equation}\label{Mittag-Leffler}
        g(k,r)= \sum_{n}\frac{a_n(r)}{k-k_n},
    \end{equation}
		where \(a_n\) are the residues of \(g\) at it's poles \(k_n\). A sufficient
		criterion for \eqref{Mittag-Leffler} is that \(g(k,r)\) falls off at
		infinity in the complex \(k-\)plane (excluding small circles around the poles) and
		the sum in \eqref{Mittag-Leffler} converges uniformly on compact sets.
    \item The integral that remains after exchange of integral and sum which is
			of the form 
    \begin{equation}
        \frac{1}{i \pi}\int_{-\infty}^\infty \frac{e^{-ik^2 t + ikr}h(k)}{k-k_n},
    \end{equation}
    may be evaluated in closed form.
\end{itemize}

In \cite{vandijk1,vandijk2,vandijk3,vandijk4}, the initial wave functions
\(\psi(r,0)\) under consideration vanishes for any \(r>R\) for some \(R>0\),
this results in growth of the fractions of \eqref{follow vd} like
\(e^{-\mathrm{Im}(k) R}\) in negative imaginary direction. The choice \(h=e^{i k
R}\) exactly cancelles this growth so that Mittag-Leffler can be applied. 

The central innovation of the current paper is a different choice of
\(h\) so as to allow for more generality:
\begin{equation}\label{Gauss-trio}
    h_\alpha(k)=e^{i\alpha  k^2}+ e^{-i \alpha k^2}+ e^{-\alpha k^2},
\end{equation}
where \(\alpha>0\) is a free parameter. 

To give an idea of which initial wave functions can be treated with this method
a quick estimate of the growth of \(C(k)\) in imaginary direction may be
helpful.  From general scattering theory we know that the eigenfunctions \(u\)
generically grow exponentially in the complex plane see \cite[equation
(12.8)]{newtonscattering}, \cite[equations (6.4.13) and (6.4.17)]{nicolathesis}
\begin{equation}\label{newton bound}
    |u(k,r)|\le \frac{c_1}{|k|} e^{|\Imi(k)|r }.
\end{equation}
If we assume that \(\psi(r,0) =\mathcal{O}\left(e^{-c_2 r^{c_3}}\right)\) for
some \(c_2>0, c_3>1\) for \(r\rightarrow \infty\) big enough a stationary phase
argument yields
\begin{equation}
	|C(k)|\le c_4 e^{(c_2 c_3)^{\frac{-1}{c_3-1}}
	(\left|\mathrm{Im}(k)\right|)^{\frac{c_3}{c_3-1}}(1-1/c_3)}.
\end{equation}
If this growth can be controlled by a Gaussian the choice \eqref{Gauss-trio}
will work for some \(\alpha>0\). Hence for any initial wave function with
\(c_3\ge2\) the method will succeed. 

The choice \eqref{Gauss-trio} results in the following expression for the wave function
\begin{align}
	\psi(r,t)=\sum_{n}a_n(r) \sum_{\gamma\in\{\alpha,\pm i\alpha\}} M(k_n,r,\gamma &+it),\label{psi expansion}
\end{align}
where \(M\) is given by
\begin{align}
    &M(k,r,\beta)=\int_{-\infty}^\infty \frac{dp}{i \pi } \frac{e^{-\beta p^2 + i r p}}{p-k}\\
    &= e^{i k r - \beta k^2} \bigg[\text{sgn}(\text{Im}(k)) + \erf\big((r+2 i k \beta )/(2 \sqrt{\beta})\big)\bigg]\notag
\end{align}
and is closely related to the Moshinsky function
\cite{moshinsky1952diffraction}. In order to apply the method one has to include
all the poles of \(g\), including the zeros of \(h_\alpha\), which are located
close to the three sets 
\begin{align}\label{Gausspoles1}
    \left\{\pm e^{-\frac{3\pi i}{8}}\sqrt{\pi(2n -1)/(\sqrt{2}\alpha)} \,\bigg| n \in\mathbb{N}\right\},\\\label{Gausspoles2}
    \left\{\pm e^{\frac{3\pi i}{8}}\sqrt{\pi(2n -1)/(\sqrt{2}\alpha)}\,\bigg| n \in\mathbb{N}\right\},\\\label{Gausspoles3}
    \left\{\pm \sqrt{\pi(2n -1)/(2\alpha)}\,\bigg| n \in\mathbb{N}\right\}.
\end{align}

\subsection{Late Time Asymptotics}
Besides the apparent use of the representation \eqref{psi expansion} to
calculate \(\psi(r,t)\) numerically with small error for early as well as for
late times it is also useful for theoretical considerations which is
illustrated by this chapter.  I will discuss the late time asymptotics of a
decaying quantum system recovering the well known result 
\begin{equation}\label{psi limit}
\psi(r,t)\xrightarrow{t\rightarrow \infty} \frac{\psi_\infty(r)}{t^{3/2}},
\end{equation}
valid vor generic wave functions. Please note, that for special initial wave
functions a different power law may apply \cite{miyamoto2004initial}.
The asymptotic behavior \eqref{psi limit} was e.g. also obtained by
\cite{vandijk3,theoryoftunneling,jensen1979spectral} for rigorous bounds see
\cite{nicolathesis,journe1991decay,schlag2007dispersive}.  The amplitude
\(\psi_\infty(r)\) will be obtained for any system to which the method of the
last section is applicable and to arbitrary accuracy in the expansion \eqref{psi
expansion}. This expression is simpler than the one of
\cite{jensen1979spectral}.  For \(t\) large relative to \(r\) and \(\alpha\), I
employ the asymptotic expression for \(\erf\) \cite[page
109-112]{olver1997asymptotics}\footnote{formulated in terms of lower incomplete
gamma function}
\begin{align}\label{erf expansion}
    \erf(\pm z)&= \pm 1-\frac{e^{-z^2}}{\sqrt{\pi}}\sum_{k=0}^M
		\frac{(-1)^k\left(\frac{1}{2}\right)_k}{ (\pm z)^{2k+1}}+\mathcal{O}(e^{-z^2} z^{-2M-3}),
\end{align}
valid for \(|\ph (\pm z)|< \frac{3}{4} \pi\) where the phase of \(k_n\) decides
which of the two expansions is to be used and \((c)_k\) is the
Pochhammer-symbol. Taking \(\gamma\in \{\alpha,\pm i \alpha\}\) and using
\eqref{erf expansion} I get
\begin{align}\label{mosh asymptotics 1}
    &\!M(k_n,r,\gamma+ i t)\approx e^{i k_n r - (it +\gamma)k_n^2}
    (\sgn(\Imi(k_n)) +\nu) \\\notag
    &- \frac{\sqrt{it + \gamma} e^{\frac{-r^2}{4(it+\gamma)}}}{\sqrt{\pi} (\frac{r}{2}+i k_n (it + \gamma))}\left(1- \frac{(it+\gamma)/2}{ \big(\frac{r}{2}+i k_n (it + \gamma)\big)^2}\right),
\end{align}
with \(\nu=-1\) for \(k_n=|k_n| e^{i\varphi}\) and \(-\frac{1}{4}\pi\le \varphi
\le\frac{3}{4}\pi\) and \(\nu=1\) otherwise.  I omitted errors of order
\(\mathcal{O}(t^{-5/2})\).  There are several remarks in order:
\begin{itemize}
	\item Except for the case \(k_n\in\mathbb{R}\) the first summand in
			\eqref{mosh asymptotics 1} is either identically zero or decays
			exponentially in time for their respective region of \(\varphi\) and may
			therefore be absorbed into the error. 
		\item For a \(k_n\in \mathbb{R}\), the first
			term is oscillatory. However, in this case \(h_\alpha(k_n)=0\) as the Jost solution \(f(-k,0)\) has no
			zeros and \(C(k)\) no poles on the real line. Hence the first term of
			\eqref{mosh asymptotics 1} vanishes in the sum over \(\gamma\in \{\alpha,
			\pm i \alpha\}\).  
	\item The complicated fractions involving \(k_n, r\) and \(t\) may be
		represented by a series in inverse powers of \(t\), which will also be a
		series in inverse powers of \(k_n\). We may then exploit the absence of a
		constant term in \eqref{Mittag-Leffler} to obtain
    \begin{align}
        0&=\sum_n \frac{a_n(r)}{k_n},\\
        \frac{C(0)}{3} \frac{f(0,r)}{ f(0,0)}&=-\sum_n \frac{a_n(r)}{k_n^2}\\
        \frac{C(0)}{3} \partial_k \left.\frac{e^{-ik r}f(-k,r)}{f(-k,0)}\right|_{k=0}&=-\sum_n \frac{a_n(r)}{k_n^3}
    \end{align}
		by inserting \(k=0\) into \eqref{Mittag-Leffler} and its derivatives. This
		yields a simpler expression for the asymptotically late wave function.
\end{itemize}
As a next step I employ the binomial series to express the
fractions in \eqref{mosh asymptotics 1} and  sum over the three possibilities of
\(\gamma\in \{\alpha,\pm i \alpha\}\) and over all poles \(k_n\), this results
in 
\begin{align}\label{psi asymptotic}
    \psi_\infty(r)
    =-\frac{C(0)}{2\sqrt{i \pi} }  \partial_k \left. \frac{f(-k,r)}{f(-k,0)}\right|_{k=0}.
\end{align}
Corrections to \eqref{psi limit} are of order
\(\mathcal{O}\left(t^{-5/2}\right)\). If \(C(0)=0\) the decay
of \(\psi(r,t)\) for large \(t\) will be faster, as observed
in \cite{miyamoto2004initial}. Higher orders can be worked out
analogously to extract an asymptotic series for \(\psi(r,t)\) in \(1/\sqrt{t}\),
which implies a series of the same kind for 
\begin{equation}\label{non-escape}
    P(t)=\int_0^\rho |\psi(r,t)|^2 dr,
\end{equation}
the probability of finding the particle inside the trapping region of length
\(\rho\). The use of the binomial series is only justified for late times, where
what ``late'' means depends on \(r\). This onset of the algebraic decay can be
estimated by equating the exponential term with slowest decay of \eqref{mosh
asymptotics 1} with the term \eqref{psi asymptotic}, yielding:
\begin{equation}\label{t_late}
    t_{\text{alg}}=\frac{-3/2}{|\Imi k_0^2|}W_{-1}\!\!\left(\frac{-|\Imi k_0^2|}{3/2^{1/3}}\left|\frac{\psi_\infty(r) \partial_{k_0}f(-k_0,0)}{k_0 C(k_0) f(-k_0,r)}\right|^{2/3}\right),
\end{equation}
where \(W_{-1}\) is the \(-1\)-branch of the  Lambert W function and \(k_0\) is
the wave vector belonging to the slowest exponential decay, i.e. the zero of
\(f(-k,0)\) in the fourth quadrant of the complex plane closest to the origin.

Besides \eqref{non-escape} the survival probability 

\begin{equation}
	S(t)=\left| \int_0^\infty \psi^*(r,0)\psi(r,t) dr\right|^2,
\end{equation}
is often used to study decaying quantum states. The long time limit of \(S(t)\)
can be found pluggin in \eqref{psi_integral} and \eqref{normalization}:

\begin{equation}
	S(t)=\left|\frac{1}{\pi} \int_{-\infty}^\infty \frac{k^2 C'(k)
	C(k)}{f(k,0)f(-k,0) h_\alpha(k)} e^{-ik^2 t} h_\alpha(k) dk\right|^2,
\end{equation}

with

\begin{equation}
	C'(k)=\int_0^\infty u(k)\psi^*(r) dr.
\end{equation}

So that Mittag-Leffler can be applied again to obtain
\begin{equation}\label{survival_mittagleffler}
S(t)=\left| \sum_{n} b_n \sum_{\gamma\in \{\alpha,\pm i \alpha\}} M(k_n,0,\gamma
+i t)\right|^2.
\end{equation}

Using the same expansion and manipulations as for the the long time limit of
\(\psi\) then yields: 

\begin{equation}\label{survival_asymptotic}
	S(t)\xrightarrow{t\rightarrow \infty}\frac{1}{t^3}~ \frac{|C(0)|^4}{4\pi (f(0,0))^4}
	,
\end{equation}
showing that \(S\) has the same power law as \(P\) and also decays faster if \(C(0)=0\).
Evaluating expansions \eqref{psi expansion} and \eqref{survival_mittagleffler}
without closed form expressions for \(f\) is difficult, this is not the case for
equations \eqref{psi asymptotic} and \eqref{survival_asymptotic}.  They may be
evaluated numerically for any potential \(V\) such that the growth conditions on
the Jost solution \(f\) and the integrals over the initial wave function \(C\)
are fulfilled.  Hence these considerations of late times may be of some use to
test non-relativistic models of alpha-decay when compared to experiment.

\section{Application to Eckart's potential}

Eckart's potential
\begin{equation}
    V(r)=\frac{A e^{r-\rho}}{\left(1+e^{r-\rho}\right)^2},
\end{equation}
where \(\rho\in\mathbb{R}\) and \(A\in\mathbb{R}\) are free parameters 
together with the initial wave function
\begin{equation}\label{psi_0}
    \psi_0(r)=\frac{2^{5/2}}{\rho^{3/2}\sqrt[4]{\pi}} r e^{-2 \left( \frac{r}{\rho}\right)^2}
\end{equation}
provides a nice example for which an exponential choice of \(h\) would not
suffice. This potential has been used to study tunneling \cite{de2022speed} from
the perspective of theoretical physics as well as from the perspective of
chemistry \cite{migliore2014biochemistry}. The associated Jost solution is
given by \cite{theoryoftunneling, eckart1930penetration}
\begin{equation}
    f(k,r)=e^{-ik r} {}_2F_1\Big(\frac{1}{2}-i \delta, \frac{1}{2}+i \delta; 1+2ik; \frac{1}{1+e^{r-\rho}}\Big),
\end{equation}
with \(\delta=\sqrt{A-1/4}\), where \({}_2F_1\) is the Gauss hypergeometric
function.

\begin{figure}[!ht] \centering \hspace*{-0.6cm}
	\includegraphics[height=7.1cm]{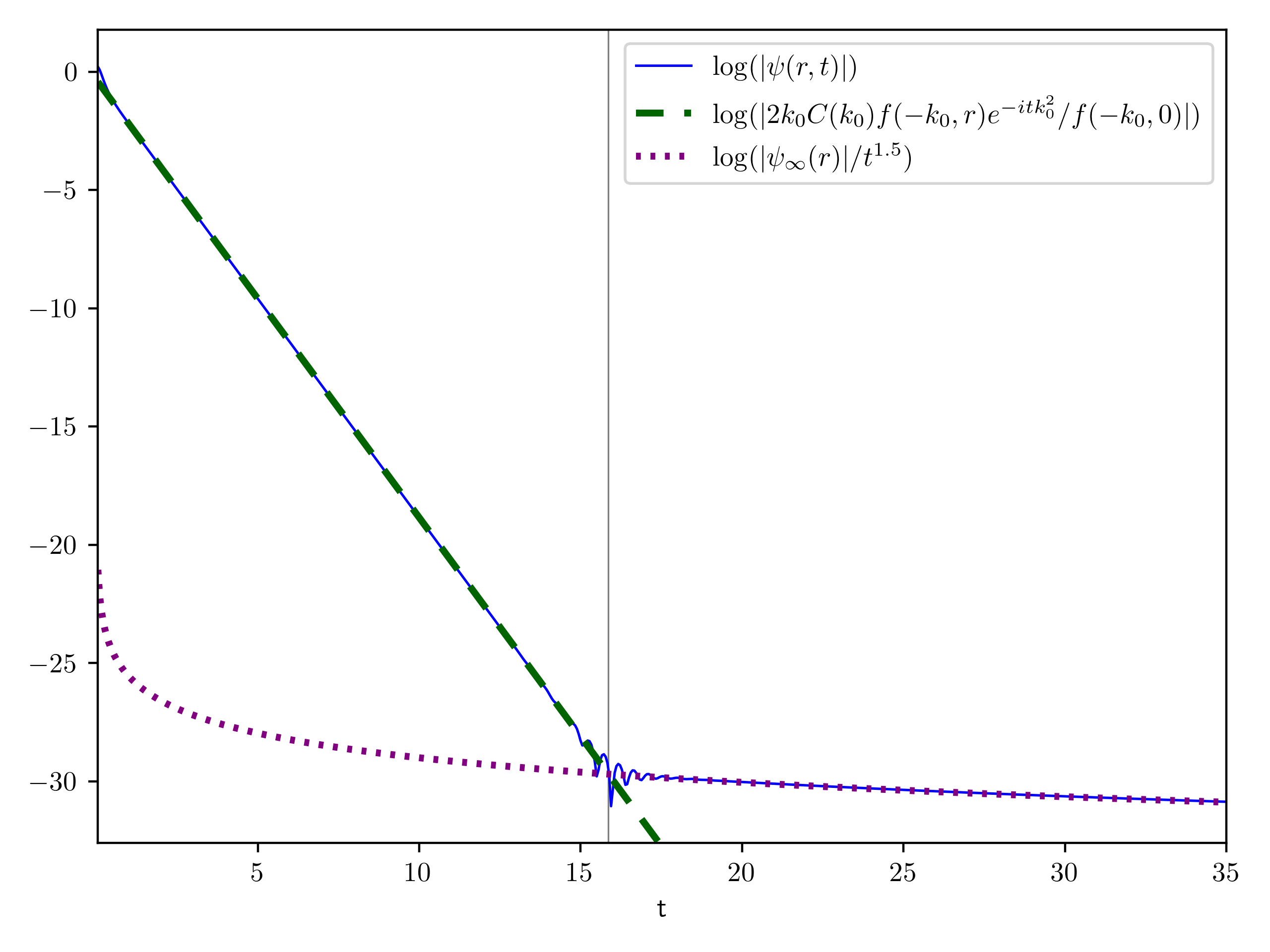} \caption{ 
		Numerical logarithmic
		comparison of \(|\psi(r,t)|\), \(|\psi_\infty(r)|/t^{1.5}\) and the slowest
		exponential decay corresponding to the zero \(k_0\) of \(f(-k,0)\) closest
		to the origin in the fourth quadrant. The thin vertical line marks
		\(t_{\text{alg}}\). The parameters are \(A=49.25, \rho=1, r=0.5\) and
		\(\alpha=1.25\). To compute \(\psi(r,t)\) all summands of \eqref{psi
		expansion} corresponding to poles \(|k_n|<100\), which results in
		approximately 18000 summands. The code that produced this figure can be found
		here: \url{https://gitlab.com/therunomask/tunneling}, it uses
		mpmath \cite{mpmath}.} \label{numerics_figure} 
\end{figure}

In this setup the Mittag-Leffler theorem implies \eqref{Mittag-Leffler}
\begin{equation}\label{repeat Mittag-Leffler}
    g(k,r)=\frac{k C(k) e^{-ikr} f(-k,r)}{h_\alpha(k) f(-k,0)}=\sum_n \frac{a_n(r)}{k-k_n}.
\end{equation}
The main reason why my choice of \(h_\alpha\) works while an exponential one
does not is as follows: 

The generalized Fourier transform of of the initial wave function, \(C(k)\),
satisfies the bound 
\begin{equation}\label{complex C bound}
    |C(k)|\le \frac{\sqrt{2} \sqrt{\rho} c_1}{|k| \sqrt[4]{\pi} }+4 c_1 \sqrt[4]{\pi}  \rho^{3/2} e^{\rho^2 |\Imi(k)|^2},
\end{equation}
which can be derived using \eqref{newton bound}.
The \(\le\) becomes \(\approx\) for \(|k|\) large enough and \(k\not\in
\mathbb{R}\). This can be compensated for by my choice of \(h_\alpha\) for
\(\alpha\) big enough, while clearly an exponential \(h\) does not suffice. 

A detailed estimate on the real line shows 
\begin{equation}
    C(k)=\mathcal{O}(k^{-6}), \quad k\rightarrow \infty
\end{equation}
implying that the sum \eqref{psi expansion} converges.  In order to evaluate
\eqref{psi expansion}, the pole structure of \(g\) should be studied. Here, in
addition to the artificial poles close to \eqref{Gausspoles1},
\eqref{Gausspoles2} and \eqref{Gausspoles3} there are poles due to the Jost
solution \(f\).  It has poles at \(k=i n/2\) for \(n\in\mathbb{N}\), however the
poles cancel in the expression for \(g\), as they appear in the numerator as
well as the denominator. The function \(f(-k,0)\) has zeros for complex \(k\),
symmetric about the imaginary axis. I conjecture the right arm of which
approaches 
\begin{equation}
    k_n=\frac{n-\frac{1}{4}+\frac{i}{2\pi}\ln\Big(e^{\pi \delta}+e^{-\pi \delta}\Big)}{i+\frac{\rho}{\pi}}
\end{equation}
for large \(n\). This conjecture is supported by numerical evidence, but I do
not have a proof yet. 

In figure \ref{numerics_figure} all of the above is used to check numerically
how accurately \(t_{\text{alg}}\) reflects the transition from exponential to
algebraic decay for the arbitrary choice \(r=1/2\). The behavior seen in figure
\ref{numerics_figure} is consistent with the analysis in \cite{vandijk3}.

\begin{figure}[!ht] \centering \hspace*{-0.6cm}
	\includegraphics[height=7.1cm]{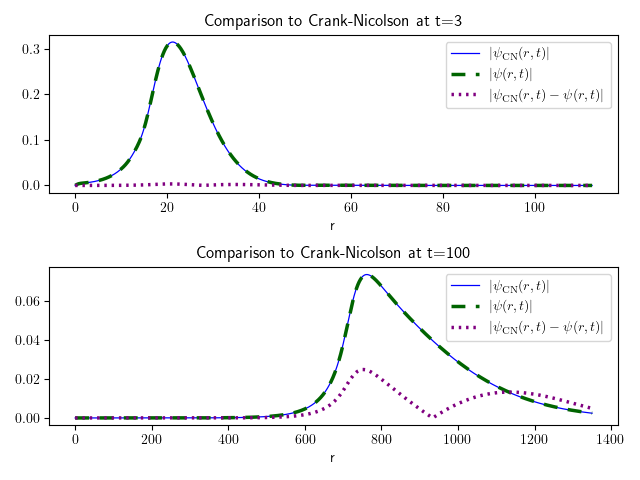} \caption{ Comparison of
		evaluation of \eqref{psi expansion} with \(A=49.25\), \(\rho=1\) and
		\(\alpha=1.25\) with the Crank-Nicolson (CN) method at \(t=3\) and
		\(t=100\). All poles with \(|k_n|\le 120\) are taken into account for
		\eqref{psi expansion}. The discretization parameters for CN are \(\Delta
		r=0.01953125\) and \(\Delta t=\Delta r^2/4\), the length of the simulation
		box is \(L=650\) for \(t=3\) and \(L=4480\) for \(t=100\). The CN method is
	implemented in Julia \cite{julia}, the code that produces this figure can also be
found here \url{https://gitlab.com/therunomask/tunneling}.} \label{CN_figure} 
\end{figure}

In figure \ref{CN_figure} the evaluation of \eqref{psi expansion} is compared
with the well established Crank-Nicolson method at an early (\(t=3\)) and
moderately late (\(t=100\)) time. One can see clearly that the two methods agree
reasonably well at early times and the absolute value of \(|\psi|\) still agrees
at \(t=100\) while substantial phase difference has accumulated leading a large
difference.

\section{Conclusion}
In the present paper I have  generalized the method of van Dijk, Nogami, and
Toyama \cite{vandijk1,vandijk2,vandijk3,vandijk4} to a large class of
non-compactly supported potentials and initial wave functions which includes
Gaussian initial states. This method was used to derive the asymptotic amplitude
\(|\psi_\infty(r)|\) as well as estimate the onset of algebraic decay
\(t_{\text{alg}}\) characteristic of large times. Finally, the method was
applied to a Gaussian wave packet  leaking through Eckart's potential  against a
hard-backwall potential barrier. 

\section{Disclaimer}
This article reflects the author's personal work and does not represent
scientific standpoints of the author's employer.

\section{Acknowledgments}
I wholeheartedly thank Christian Bild for meticulously checking my calculations
as well as Siddhant Das for helpful discussions. 

\nocite{*}

\bibliography{apssamp.bib}% Produces the bibliography via BibTeX.

\end{document}